\documentclass[prl,twocolumn,aps,showpacs,superscriptaddress]{revtex4}
\usepackage{graphics}
\usepackage{epsfig}
\usepackage{amsmath}
\usepackage{amssymb}
\usepackage{bm}
\usepackage{color}

\begin{document}
\title{Relaxations and rheology near jamming}

\author{Brian P.~Tighe}
\altaffiliation[Current address: ] {Delft University of Technology, Process \& Energy Laboratory, Leeghwaterstraat 44, 2628 CA Delft, The Netherlands}
\affiliation{Instituut-Lorentz, Universiteit Leiden, Postbus 9506, 2300 RA Leiden, The Netherlands}

\date{\today}

\begin{abstract}
We determine 
the form of the complex shear modulus $G^*$ in soft sphere packings near jamming.
Viscoelastic response at finite frequency is closely tied to a packing's intrinsic relaxational modes, which are distinct from the vibrational modes of undamped packings. 
We demonstrate and explain the appearance of an anomalous excess of slowly relaxing modes near jamming, reflected in a diverging relaxational density of states. 
From the density of states, we derive the dependence of $G^*$ on frequency and distance to the jamming transition, which is confirmed by numerics.

\end{abstract}

\maketitle

\begin{figure}[tbp] 
\centering
\includegraphics[clip,width=0.9\linewidth]{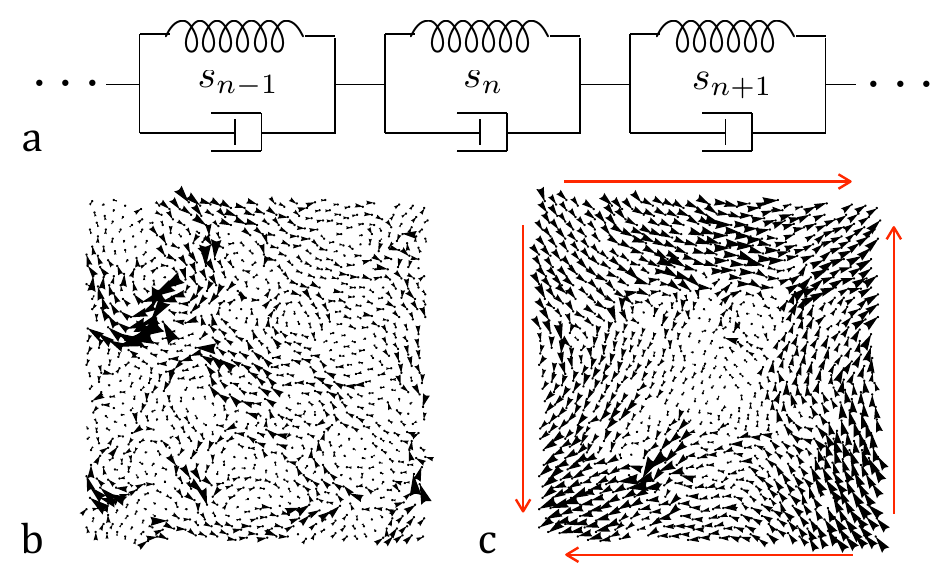}
\caption{ 
(a) Kelvin-Voigt viscoelastic elements in series. Elements are made of a spring with stiffness $G_n$ in parallel with a dashpot with coefficient $\eta_n$; their characterisic relaxation rate is $s_n = G_n/\eta_n$. (b) Disk displacements of a relaxational mode with rate $s_n = 0.006$ in a packing with excess coordination $\Delta z = 0.013$; the disks are not shown. (c) Shear response for the same packing driven at $\omega = 0.006$.
}
\label{fig:summary}
\end{figure}
When the bubbles or droplets that make up foams or emulsions are packed closely together, they jam into a mechanically rigid state
\cite{princen86,bolton90,vanhecke10,clusel09,katgert10b}. 
Near jamming they form amorphous packings of repulsive athermal spheres \cite{bolton90}, and in recent years the linear quasistatic response of jammed packings has been 
exhaustively 
mapped out. Proximity to (un)jamming, a nonequilibrium critical point, organizes their elasticity, and in simulations their moduli scale with distance to the transition \cite{vanhecke10}. Experimental evidence for these scalings, however, remains elusive. Rheology may prove a better test bed for the jamming paradigm, as there is already a wealth of available data \cite{katgert09,mason97,liu96,gopal03}. While there is growing numerical and theoretical evidence that jamming also organizes the rheology of soft sphere packings, details remain controversial \cite{durian97,olsson07,hatano09}.

Prior rheological studies of the jamming transition have focused on steady flow  \cite{katgert09,durian97,olsson07,hatano09}. Surprisingly, despite its practical and fundamental significance, little is known about the oscillatory rheology of soft spheres near jamming.
The complex shear modulus $G^*(\omega) = G'(\omega) + \imath G''(\omega)$, composed of the storage modulus $G'$ and loss modulus $G''$, describes the linear response to oscillatory driving at frequency $\omega$. It is a fundamental material characterization, encoding the response to shear on all time scales, including the quasistatic (QS) shear modulus $G_0$ in the zero frequency limit. 
Experiments provide evidence as to the form of $G^*$: liquid and organic foams, emulsions, and microgel suspensions are all shear thinning, $G^* \sim (\imath \omega)^{\Delta}$ with $\Delta \approx 0.5$, over one to several decades in frequency \cite{liu96,gopal03}.

Whereas undamped packings have vibrational normal modes, modes in overdamped packings are {\em relaxational}. In this Letter we relate relaxational modes and rheology to determine and explain the dependence of $G^*$ on frequency and distance to jamming. $G^*$ displays dynamical critical scaling; shear thinning is a critical effect and the shear thinning regime extends to zero frequency at the jamming transition. We calculate all critical exponents, including $\Delta = 1/2$. This makes $\Delta$ one of the few experimentally observed jamming exponents. 

The relationship between relaxations and rheology has a simple schematic expression: a packing's finite frequency linear response can be mapped to a series circuit of Kelvin-Voigt viscoelastic elements (Fig.~\ref{fig:summary}a). Each element has a characteristic relaxation rate $s$ equal to the eigenrate of one of the packing's relaxational modes (Fig.~\ref{fig:summary}b). Thus the response to driving (Fig.~\ref{fig:summary}c) is governed by the distribution of relaxation rates, i.e.~the density of states $D(s)$, which is related to, but different from, the vibrational density of states $D(\Omega)$. Here we will determine $D(s)$ for the first time, use scaling arguments to identify and explain its characteristic features, and from these
predict the form of $G^*(\omega)$.

{\em Model system.}--- 
Starting from numerically generated static packings, we study their linear response: interactions are linearized about the reference state and deformations are calculated without allowing for rearrangements of the contact network. Our jammed packings in $d=2$ dimensions are comprised of weakly polydisperse disks in a $L \times L$ unit cell generated using the molecular dynamics protocol of Ref.~\cite{somfai05}. 
Packings are characterized by their mean number of contacts per particle, $z$. Static packings at the jamming transition are isostatic, $z = z_c = 2d$, hence the excess contact number $\Delta z = z - z_c$ serves as a measure of distance to jamming \cite{vanhecke10,footnote0}.  

We impose the dynamics of the `full bubble model' of Durian \cite{durian97,footnoteMF}, in which foam bubbles are modeled as disks. Contacting disks interact via elastic and repulsive forces $f^{\rm el} = k \delta$ that are linear in the overlap $\delta$, with a spring constant $k$. Touching disks also experience a viscous force ${\vec f}^{\rm visc} = b \, \Delta \vec v$, with damping coefficient $b$, opposing their relative velocity evaluated at the contact, $\Delta \vec v$. The dynamics are overdamped, 
so that forces and torques balance on each bubble at every instant. 
We report stresses in units of $k$ and times in units of $b/k$.

To describe a shear deformation, which involves motion of the disks and distortion of the unit cell, requires $3N+1$ degrees of freedom: the $3N$ displacements and rotations of the disks, $\lbrace u_{x,i}, u_{y,i}, \theta_i\rbrace$, plus the magnitude $\gamma L$ of the pure shear displacement applied to the lattice vectors of the unit cell. Equations of motion governing the disks' motions {\em and} the unit cell's deformation are most easily written in a Lagrangian formalism. To implement this we note that the elastic potential energy $V$ is
\begin{equation}
V = \frac{1}{2}\sum_{\langle ij 
\rangle} \left[
k (\Delta u^\parallel_{ij})^2 
- \frac{f^0_{ij}}{\Delta r^0_{ij}} 
(\Delta u^\perp_{ij})^2 
\right] \,,
\label{eqn:potential}
\end{equation}
where $\Delta u^\parallel$ and $\Delta u^\perp$ label normal and tangential relative displacements, possibly across periodic boundaries.  $f^0$ and $\Delta r^0$ are the force and separation between disks in the reference state. Similarly, for the viscous forces described above, the Rayleigh dissipation function $R$ is \cite{goldstein}
\begin{equation}
R =\frac{1}{2}b\sum_{\langle ij 
\rangle} \left[
(\Delta {\dot u}^\parallel_{ij})^2 +
(\Delta {\dot u}^\perp_{ij} 
- \rho_i {\dot \theta_i} - \rho_j {\dot \theta_j})^2 
\right] \,.
\label{eqn:dissfn}
\end{equation}
The $\lbrace \rho_i \rbrace$ are disk radii. The disk rotations enter because relative velocities are evaluated at the contact. The forms of $V$ and $R$ will be needed below to extract the scaling of the relaxational density of states.

The Lagrangian equations of motion are
\begin{equation}
{\hat K}|{ q}(s)\rangle  
+ s{\hat B}|{ q}(s)\rangle 
= \sigma(s)L|\hat \gamma\rangle  \,,
\label{eqn:eom}
\end{equation}
where the matrices are $K_{mn} = \partial^2 V/\partial q_m \partial q_n$ and $B_{mn} = \partial^2 R/\partial \dot q_m \partial \dot q_n$, $|\hat \gamma \rangle$ is a unit vector along the strain coordinate, and $\sigma$ is the shear stress. We have collected all $3N+1$ degrees of freedom in a vector $|q(t)\rangle$ and applied a Laplace transformation; transformed quantities depend on the independent variable $s$, which has units of inverse time and may be thought of as a relaxation rate.

{\em Relaxational density of states.}--- 
\begin{figure}[tbp] 
\centering
\includegraphics[clip,width=0.95\linewidth]{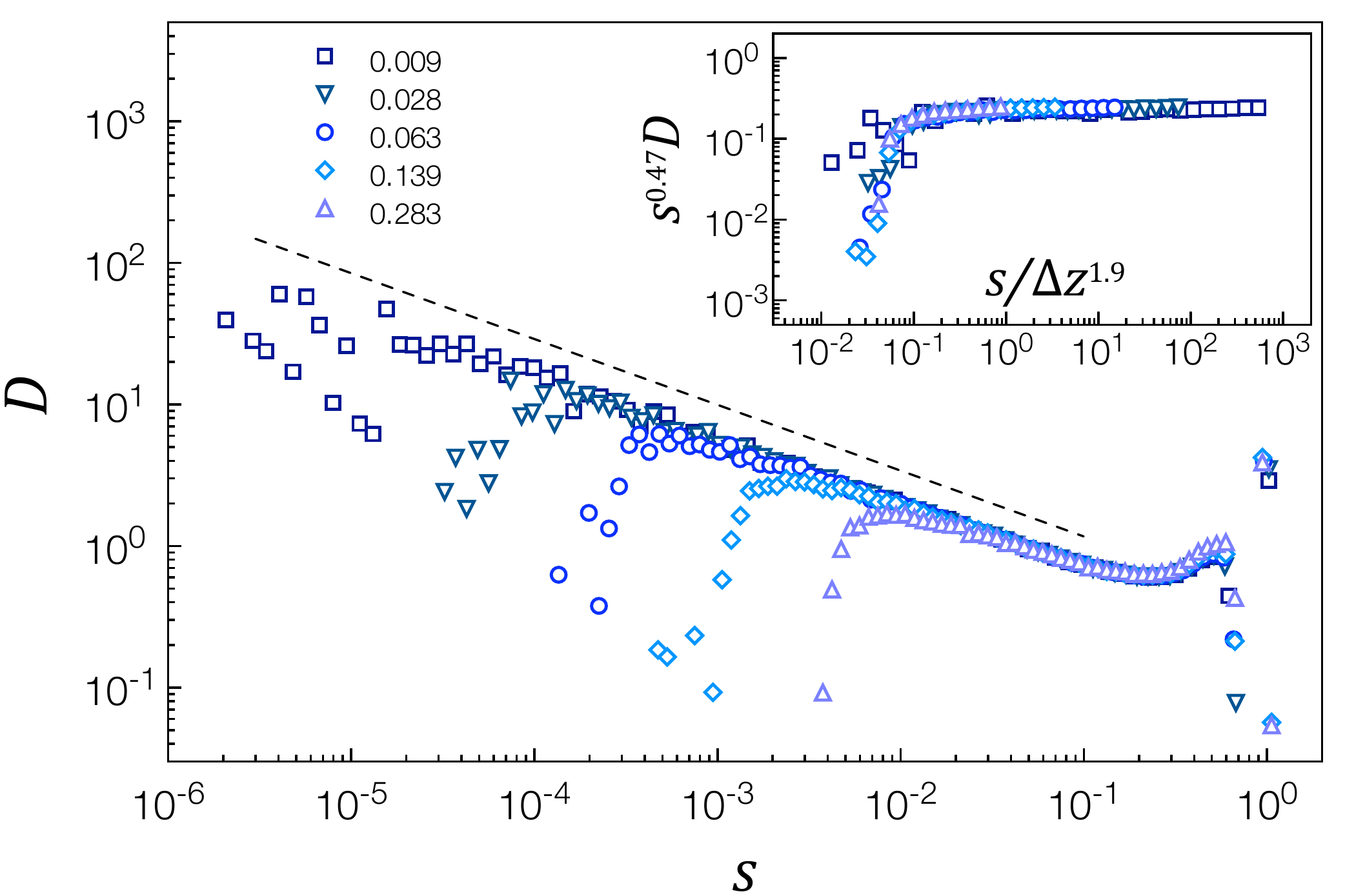}
\caption{ Relaxational density of states. As the excess coordination number $\Delta z$ (legend) vanishes, $D(s)$ develops a power law divergence. The dashed curve $\sim 1/s^{0.47}$. Inset: The low $s$ data is collapsed by rescaling $s$ with the crossover rate $s^* \sim \Delta z^{1.9}$. Exponents are best fits (see text).}
\label{fig:dos}
\end{figure}
Before considering driving, we study the system's relaxational modes and rates. For $\sigma = 0$, Eq.~(\ref{eqn:eom}) is a generalized eigenvalue equation; its eigenvalues and -vectors are the relaxational rates and modes;
Fig.~\ref{fig:summary}b gives an example.
When deformed along the mode $|s_n\rangle$, the system relaxes exponentially to equilibrium with a rate given by the eigenvalue $s_n \le 0$
\cite{footnote2}.  We use $s$ to refer to $|s|$ whenever no ambiguity results.

Fig.~\ref{fig:dos} displays the relaxational density of states $D(s)$, the distribution of rates, averaged over approximately 50 packings of $N=1024$ disks for multiple values of $\Delta z$.
Several features stand out. First, close to jamming ($\Delta z\rightarrow 0$), $D(s)$ develops a power law divergence:  packings at the transition possess many {\em slowly} relaxing modes. Second, the divergence is cut off at a crossover rate $s^*$ that vanishes with $\Delta z$, so that, as packings are prepared closer and closer to the transition, the weight of slow modes grows. Finally, the fastest relaxations are on the order of the bare rate $k/b = 1$. 

{\em Scaling of $D(s)$.---} 
To explain the characteristic features of $D(s)$, we generalize the ``cutting argument'' of Wyart, Nagel, and Witten (WNW) to overdamped dynamics. In seminal work, WNW introduced this variational argument to explain the low frequency plateau in the {\em vibrational} density of states of undamped packings, $D(\Omega)$ \cite{wyart05}. Here we sketch relevant details.

The WNW argument involves constructing a set of trial modes and estimating their frequencies, from which the scaling of $D(\Omega)$ can be inferred. The trial modes are made by first ``cutting'' contacts at the boundary of patches of size $\ell$ -- this introduces floppy modes, zero frequency collective modes that involve no relative normal motions between disks -- and then ``stitching up'' the cut with a sinusoidal envelope with wavenumber $q \sim 1/\ell$. Because of the sinusoid there are small relative normal motions in the trial mode. Trial modes can be constructed for wavenumbers $q \gtrsim q^* \sim \Delta z$, and their probability density is flat, $D(q) \sim q^0$ for $q \gtrsim q^*$.

To generalize these results to damped dynamics, there are two key observations. First, we anticipate that, for sufficiently low frequencies, disk motions smoothly approach their quasistatic values. Thus for long time scales the trial modes are also good trial modes for the overdamped dynamics. Second, $D(s)$ can be related to $D(q)$ by inferring the ``dispersion relation'', i.e.~how $s$ scales with $q$. If $s \sim q^{1/\nu}$, then $D(s) \sim q^0 |{\rm d}q/{\rm d}s| \sim s^{\nu - 1}$.

To estimate the overdamped dispersion relation, we assume that a trial mode $|q_{\rm WNW}\rangle$ is an approximate solution to the homogeneous equation of motion, $\hat K |q_{\rm WNW} \rangle + s \hat B |q_{\rm WNW} \rangle \approx 0$. Taking an inner product and bearing in mind Eqs.~(\ref{eqn:potential}) and (\ref{eqn:dissfn}) for $V$ and $R$, at the transition ($z=z_c$) one finds
\begin{equation}
s \sim 
\frac{k}{b}
\frac{
(\Delta  u^\parallel_{\rm WNW})^2 
}{
(\Delta  u^\parallel_{\rm WNW})^2 + 
(\Delta  u^\perp_{\rm WNW})^2 
}\,,
\label{eqn:viscous}
\end{equation}
where for scaling we refer to typical (relative) displacements.
Normal motions in a trial mode come from the sinusoidal modulation: $\Delta  u_{\rm WNW}^\parallel  \sim \partial u_{\rm WNW} \sim q \, u_{\rm WNW}$. In contrast, tangential motions are predominantly contributed by the floppy mode -- they persist if $q$ is sent to zero -- so to leading order they are independent of $q$, $\Delta  u_{\rm WNW}^\perp \sim u_{\rm WNW}$. In the limit of small $q$, then, the dispersion relation is quadratic, $s \sim q^2$. Hence $\nu = 1/2$, and at jamming the density of states diverges as
\begin{equation}
 D(s) \sim s^{-1/2} \,.
\label{eqn:squareroot}
\end{equation}
For $z>z_c$, the dispersion relation remains quadratic for $s \gtrsim s^* := (q^*)^2$, or 
\begin{equation}
s^*  \sim \Delta z^{2} \,,
\label{eqn:taustar}
\end{equation}
so Eq.~(\ref{eqn:squareroot}) continues to hold above $s^*$. Eqs.~(\ref{eqn:squareroot}) and (\ref{eqn:taustar}) are our first main result. 

The predicted divergence and crossover scaling in $D(s)$ can be tested by plotting $s^{\Delta}D(s)$ versus $s/\Delta z^{\lambda}$, as shown in the inset of Fig.~\ref{fig:dos}. Eqs.~(\ref{eqn:squareroot}) and (\ref{eqn:taustar}) predict that the crossovers at low $s$ will collapse for $\lambda = 2$, and that above the crossover the curves will be flat for $\Delta = 1/2$. We find the best collapse and plateau for $\lambda = 1.9(1)$ and $\Delta = 0.47(5)$, in good agreement with our predictions.  

The vanishing rate $s^*$ implies a diverging time scale $\tau^* := 1/s^* \sim 1/\Delta z^{2}$.
We stress that $\tau^*$ is different from the diverging time scale $ \sim 1/\Delta z$ observed in undamped packings \cite{wyart05}. 
These two time scales come from different limiting cases of dynamics with both damping and inertia; when inertial effects dominate damping the dispersion relation is linear rather than quadratic.

{\em Shear response.}---
Near jamming there is an abundance of slow relaxational modes. These must influence the response to shear (Fig.~\ref{fig:summary}c). We now show that the anomalous modes are responsible for shear thinning at frequencies $\omega \gtrsim s^*$, while for $\omega \lesssim s^*$ the response is quasistatic.

\begin{figure}[tbp] 
\centering
\includegraphics[clip,width=0.95\linewidth]{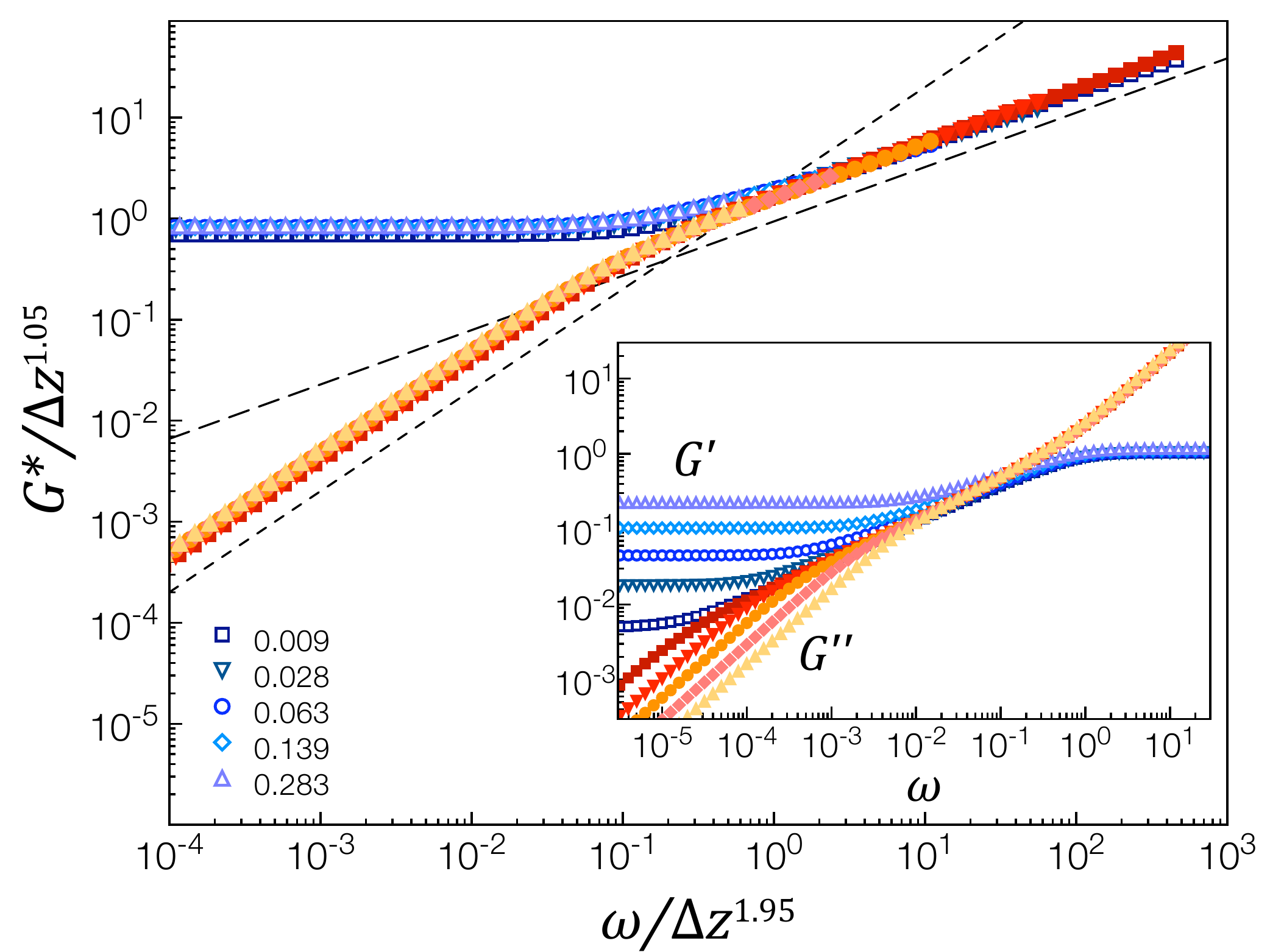}
\caption{ Data collapse of the storage modulus $G'$ (open symbols) and loss modulus $G''$ (filled symbols) plotted for $\omega \leq 0.07$ and varying distance to jamming $\Delta z$ (legend). The long (short) dashed curve indicates $\omega^{0.54}$ ($\omega$) scaling. Exponents are best fits (see text). Inset: $G^*$ prior to rescaling.}
\label{fig:response}
\end{figure}

For a sinusoidal strain with frequency $\omega$, the shear stress is $\sigma(t) = G^*(\omega) \gamma(t)$.
Numerical results for $G^*$ on approach to jamming are plotted in the inset of Fig.~\ref{fig:response}  
\cite{footnote1}. At high frequency $\omega \gtrsim 1$, loss dominates storage and both moduli reflect their microscopic counterparts: $G^{\prime\prime}/\omega \sim G^{\prime} \sim {\cal O}(1)$. This is because the packing cannot relax faster than the bare rate $k/b$ and there is no inertia, hence high frequency deformations are affine and viscous stress dominates. Elasticity dominates for low frequencies, with the quasistatic modulus $G_0 := \lim_{\omega \rightarrow 0} G^{\prime}$ diminishing and the dynamic viscosity $\eta_0 := \lim_{\omega \rightarrow 0} G^{\prime\prime}/\omega$ growing on approach to jamming.  
Between these two extremes is a range of frequencies over which $G'$ and $G''$ grow with sub-linear dependence on frequency, i.e.~shear thinning; this range grows as jamming is approached.

{\em Scaling of $G^*(\omega)$.}---
By expanding $|q(s)\rangle$ in the relaxational modes and using the fact that the modes are $\hat B$-orthogonal, $\langle s_m | \hat B | s_n \rangle = 0$ for $m \neq n$, Eq.~(\ref{eqn:eom}) can be inverted to give an exact expression for $G^*$:
\begin{equation}
\frac{1}{G^*(\omega)} = \sum_{\lbrace |s_n| > 0 \rbrace} 
\frac{1}{G^*_{n}(\omega)}  \,,
\label{eqn:responsefunction}
\end{equation}
with $G^*_n(\omega) = (|s_n| + \imath\omega)\langle s_n | \hat B | s_n \rangle        /|\langle s_n | \hat \gamma \rangle |^2$.  
Such a reciprocal sum describes the modulus of a series circuit of viscoelastic elements. Further, the form of $G^*_n(\omega)$ is identical to the shear modulus of a Kelvin-Voigt element with rate constant $|s_n|$; each mode acts like a Kelvin-Voigt element in the circuit (Fig.~\ref{fig:summary}a). 
We stress that Eq.~(\ref{eqn:responsefunction}) is not a model postulated {\em ad hoc}, but follows directly from inverting the equations of motion. It makes clear that linear rheology is controlled by the distribution of relaxation rates, $D(s)$.%

The scaling of $G^*$ can be extracted from Eq.~(\ref{eqn:responsefunction}). Treating each mode's shear component $\langle s_n | \hat \gamma \rangle$ as a random variable independent of $s_n$, 
near jamming one finds
\begin{equation}
 \frac{1}{G^*(\omega)} \gtrsim \int_{s^*}^{1} 
 \frac{s^{-1/2}\,{\rm d}s}{s+ \imath \omega}
 \,.
\label{eqn:asymptotics}
\end{equation}
Asymptotic analysis of the integral gives
\begin{equation}
G^\prime
\sim \left \lbrace
  \begin{array}{c}
  \Delta z  \\
  \omega^{1/2}  \\
  1 
  \end{array} \right.
  \mbox{and }\,
G^{\prime\prime} 
\sim \left \lbrace
  \begin{array}{cccccl}
  \omega / \Delta z &  \: \omega \!  & \! \lesssim \! & \! s^* \!  & \!  \! & \!  \\
  \omega^{1/2} &  \: s^* \! & \! \lesssim \! & \! \omega \! & \! \lesssim \! & \! 1 \\
  \omega &   \:  & & 1 & \! \lesssim \! & \omega \,.
  \end{array} \right.  
\label{eqn:Gp}
\end{equation}
This is our second main result.

Eq.~(\ref{eqn:Gp}) provides (i) an analytical prediction for the scaling of the quasistatic shear modulus $G_0 \sim \Delta z$, consistent with numerical and analytical modeling in the QS limit \cite{vanhecke10,wyart05b}.  It predicts (ii) a diverging viscosity $\eta_0 \sim 1/\Delta z$,
which has not previously been predicted or measured.
It shows (iii) that $G'$ deviates qualitatively from the QS limit for $\omega \gtrsim s^*$. Thus the QS approximation is only reasonable for time scales much longer than  $\tau^* \sim 1/\Delta z^2$, which diverges at jamming. 
Finally, (iv), Eq.~(\ref{eqn:Gp}) predicts that the storage and loss moduli display a shear thinning regime scaling as $\omega^{\Delta}$, with $\Delta = 1/2$ being the same exponent that governs the divergence of $D(s)$.
The regime extends to frequencies as low as $\omega \sim s^*$, and hence to zero frequency at the jamming transition. Such scaling is therefore a critical effect, and the critical jammed state is a shear thinning complex fluid. 

Eq.~(\ref{eqn:Gp}) also predicts that plotting the rescaled complex modulus $G^*/\Delta z^\mu$  and rescaled frequency $\omega/\Delta z^{\lambda}$ will produce data collapse for $\mu = 1$ and, again,  $\lambda = 2$. Indeed we find excellent collapse for $\mu = 1.05(5)$ and $\lambda = 1.95(5)$ (Fig.~\ref{fig:response}), in good agreement with the predicted values. The loss modulus is linear for low rescaled frequency (short dashed curve), as predicted, while for larger values both $G'$ and $G''$ approach a power law scaling $G^* \sim (\imath \omega)^\Delta$ (long dashed curve); the value $\Delta = \mu / \lambda = 0.54(4)$ again agrees well with its predicted value of $1/2$. As Eq.~(\ref{eqn:Gp})  also captures the trivial high frequency scaling, its predictions are fully confirmed. While our numerics are restricted to two dimensions, the predicted exponents are independent of $d$, as are all known jamming exponents \cite{vanhecke10}.

{\em Outlook.}--- 
We  have identified and explained the divergence and vanishing crossover in the relaxational density of states. We have also shown that slow relaxational modes determine the form of the complex shear modulus near jamming, including the quasistatic shear modulus $G_0$ and dynamic viscosity $\eta_0$, the critical shear thinning regime, and the diverging relaxation time $\tau^*$ governing their crossover. 

The predicted QS scaling has yet to be observed experimentally; low frequency rheology is complicated by aging processes such as coarsening, which  are not modeled here \cite{gopal03}.
By contrast, the anomalous shear thinning regime has already been observed  \cite{liu96,gopal03}. 
While prior work by Liu et al.~also predicts $\omega^{1/2}$ scaling  \cite{liu96}, it invokes the fluctuation-dissipation theorem and does not explain the connection to other scaling regimes or to jamming. Molecular dynamics simulations {\em below} jamming also find $\Delta \approx 0.5$, consistent with our prediction \cite{hatano09}. 

We have demonstrated that collective effects can give rise to anomalous $\omega^{1/2}$ scaling in a system where microscopic viscous forces are linear. One might also look for sources of nonlinearity in the microscopic interactions, but we know of no such mechanism common to liquid and organic foams, emulsions, and microgel suspensions, all of which display the same shear thinning rheology \cite{gopal03}. We therefore conclude that $\Delta = 1/2$ is one of the few  experimentally observed jamming exponents \cite{katgert10b}.

\begin{acknowledgments}
It is a pleasure to thank S.~Dagois-Bohy, M.~van Hecke, W.~van Saarloos, and V.~Vitelli for discussions and the Aspen Center for Physics for hospitality. This work was supported by the Dutch Organization for Scientific Research (NWO).
\end{acknowledgments}


\begin{thebibliography}{16}
\bibitem{footnote0}{
$\Delta z$ is related to the distance to the critical packing fraction $\Delta \phi = \phi - \phi_c$ and pressure $p$ via the scaling relations $\Delta z \sim \Delta \phi^{1/2}$ and, for spring-like forces, $p \sim \Delta \phi$ \cite{vanhecke10}.}
\bibitem{footnote2}{
Invariance under global translations and local rotations contributes $N+2$ zero modes, not plotted in Fig.~\ref{fig:dos}.}
\bibitem{footnote1}{
Packings can be unstable to shear at certain orientations. We remove the prestress term in $V$ to enhance stability. 
}
\end{thebibliography}
\end{document}